# SnB Collaborative Visualization

**Dave Pape[1], Amin Ghadersohi[2], Josephine Anstey[1], Amit Makwana[3]**

(1) Department of Media Study
(2) Department of Computer Science and Engineering
(3) Department of Mechanical Engineering
University at Buffalo, Buffalo, NY 14260, USA
E-mail: depape@buffalo.edu

Abstract
*We describe a system for visualization and editing of data in a computational chemistry environment. The system is a collaborative tool allowing researchers using virtual reality and/or desktop computer displays to work together on results of the Shake-and-Bake structure determination application.*

## 1  Introduction

Analysis and visualization of protein structures is an important field of work in chemistry research, relevant to many application areas. Shake-and-Bake (Miller et al, 2004) is a state-of-the-art program for molecular crystal structure determination from X-ray crystallographic data. The SnB package, which has been distributed to more than 500 laboratories worldwide, was originally released in the mid-1990s with a primitive, single-user back-end visualization package.

Due to the distributed fashion in which the structure determination community can operate, there is a significant need for an interactive collaboratory that will work simultaneously in a variety of settings, from high-end immersive VR environments to desktop PCs. While simple distributed molecular visualization systems have existed for some time, none of these packages provide the functionality required to edit structures in a collaborative, real-time fashion.

Another important aspect of working with Shake-and-Bake is that the output structure determination program is often not precise. In a typical application scenario, an expert user (i.e. a chemist) must take the program output and edit it, using domain knowledge and knowledge of the particular structure being examined to complete the work – e.g. adding bonds where there should be bonds, removing others that could not exist, and labelling the types of the different atoms. Standard molecular visualization tools do not provide all the capabilities needed in this case.

Ideally, we wished to make use of a virtual reality environment, to explore the advantages of the first-person perspective and direct physical manipulation of virtual objects when viewing and editing the molecular structures. However, not all chemistry research labs will have ready access to VR systems; to make the system more widely usable, particularly for collaboration between multiple labs, we needed to develop a cross-platform tool, one that supports both VR displays and ordinary desktop computers.

## 2  Related Work

RasMol (Sayle and Milner-White 1995), one of the earliest tools for protein visualization, offers a fast C based engine that allows viewing of protein structures. RasMol is now being developed under the name of Protein Explorer. However, RasMol does not offer any editing or collaboration features.

The Molecular Collaborative Environment (MICE) (Bourne et al. 1998), a Java™ based project that aims at ease of use mixed with useful features, offers some interactive collaborative features. MICE defines a Molecular Scene Description Language that allows molecular structures to be stored in a rational database (a molecular scene gallery) and queried. Structures can be retrieved from the gallery, rendered in Virtual Reality Modeling Language and displayed in a WebView browser modified to support the Virtual Reality Behavior System protocol.

Another tool is The Chimera Collaboratory Extension (UCSF 2004), which is an application add-on to the Chimera molecular modeling system. The Collaboratory provides functionality for synchronizing an interactive three-dimensional molecular modeling session in real-time between multiple users

Despite the fact that most molecular visualization programs provide analysis and viewing functionality, they do not provide a collaborative mechanism for editing structures.

## 3  Shake-and-Bake Visualizer

Our visualization system, SnB Visualizer, is built using a simple scene graph structure implemented in C++ and OpenGL. By using libraries widely available on multiple platforms, we were able to achieve cross-platform compatibility across Windows, MacOSX, and UNIX. Moreover, by making an abstraction above the basic input/output system, we were able to achieve concurrent compatibility with most windowing toolkits such as WIN32, the OpenGL Utility Toolkit (GLUT), as well as multi-pipe CAVEs™ (Cruz-Neira et al, 1993). In developing the Visualizer, our primary VR system was a Linux-based, single-screen VR display; the system comprises a large, rear-projected passive stereo screen, 6 degree-of-freedom electromagnetic tracker with sensors on the head and two hands, and wand control device (Pape et al, 2002). This type of low-cost VR system is one that could reasonably be deployed in many chemistry research labs. Figure 1 shows the VR version of the SnB Visualizer in use.

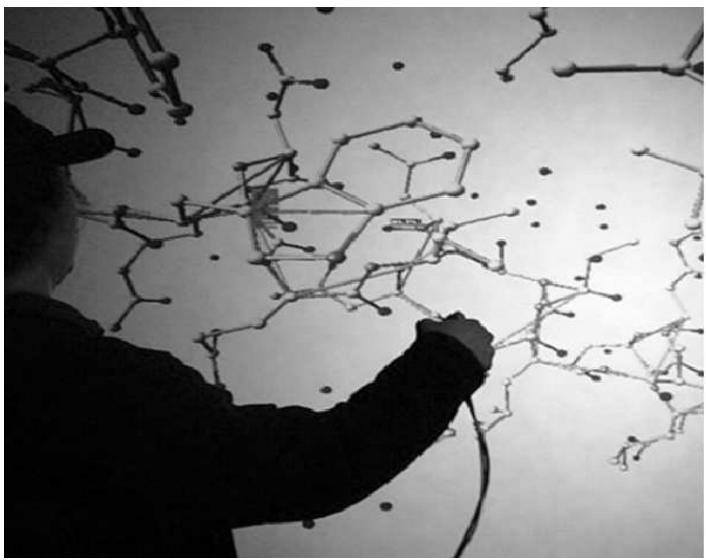

*Figure 1. Virtual reality interface for SnB visualizer.*

In our system, the user interaction is provided through a Java-style user interface library created on top of our simple scene graph. This provides user interaction via mouse and keyboard on a desktop system and via a wand in the CAVE environment. Coupled with a networked database built on top of TCP/IP, the SnB Visualizer serves as a collaborative viewing and editing tool for 3D protein structures (figure 2).

The SnB Visualizer provides the functionality to distribute multiple datasets across a network so that multiple users on different remote clients can view and interactively modify any part of the scene. Specifically, it can load multiple molecules, allowing different users to edit any one of them at any time. In addition, it serves as a real-time monitor for any algorithm with output that is in the form of a graph composed of nodes (vertices) and links (edges). For example, the Shake-and-Bake program, which outputs 3D protein structures, can have many instances running at the same time. Each instance outputs a file that contains atom coordinates and bond information. All the output files can be monitored by one instance of SnB Visualizer running as a server.

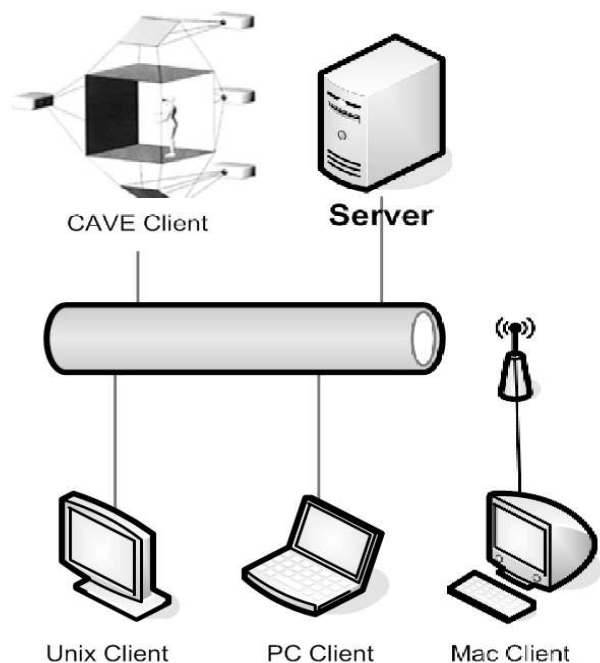
*Figure 2. Collaborative cross-environment visualization.*

### 3.1 Interaction

In this section, we describe the techniques we developed to provide real-time interaction with the molecules. Our current implementation provides support for the GLUT and CAVE environments. The ultimate goal is to be able to select different objects in the scene and modify them (e.g., atoms and bonds). GLUT provides a means for interaction through a mouse and keyboard, and CAVE uses a wand and joystick. A wand is basically the handle for a projected ray or line segment. In a virtual environment, we can use a wand to project a line that can intersect objects in the scene. Using collision detection techniques, atoms and bonds can be tested for intersection with the line projected from the wand. We have created an abstract Wand object that gets its position and orientation from the CAVE engine. In cases where a mouse is used, the coordinates of the mouse pointer is used to project a ray from the center of the screen to the position of the mouse. This, in effect, acts like a wand that one may be holding in front of them in an actual CAVE. The signals from the joystick can be interpreted the same way as the keyboard buttons with an option to indicate amount of displacement since the joystick may be pressure sensitive.

Like many other graphics environments, GLUT and the CAVE library (VRCO 2004) are callback based. For example, we provide GLUT or the CAVElib with a function pointer to a per-frame draw and update function, as well as an initialization function to configure the required variables. Based on the types of input devices (keyboard, mouse, wand, tracker, joystick, etc.) and event systems (polling, interrupt driven) used, other callback functions need to be implemented. For example, GLUT requires callback function for mouse, keyboard, joystick, window and other events, while the CAVElib leaves it up to the programmer to detect changes in input values using predefined global variables. Other systems such as Microsoft Foundation Classes (MFC) and Cocoa (Apple Computer, 2005) also use similar mechanisms for notification of input change.

We have created an abstract IO class to act as an interface for extending the visualizer to operate with other windowing toolkits, as mentioned above. For example, the Molecule class, derived from Graph, is managed by MoleculeViewer, which receives its input from IO and its derivatives. Furthermore, collision detection between the wand and graphical elements is inherited from a set of collidable classes such as LineSegment, Sphere and Cylinder.

### 3.2 User Interface

There are many user interface toolkits available. However, many of them are tailored to a specific environment. For the purposes of the visualizer, we needed user interface components that are navigable with CAVE using the wand, as well as on desktops by way of mouse and keyboard. We designed a set of classes for this purpose similar to the Java AWT/Swing classes. In a regular desktop environment, the buttons, labels, menus and other user interface components are rendered to the screen as a heads-up-display using a second orthographic camera, allowing for direct use of mouse coordinates for interfacing with these components. However, using the CAVE, these user interface components are displayed as objects in 3D space, just like the molecules, intersectable by the wand. This allows us to use the wand buttons to perform operations on these components when they are intersected.

## 4 Conclusion

In this paper, we have presented a cross-platform, multi-environment visualization system called the *SnB Visualizer*. Our implementation is object-oriented and fully modularized. The system employs various techniques such as wand simulation and ray-tracing to facilitate interaction with the 3D models in VR or regular desktop environments. The networked database module can be used in any program that requires real-time networking of data. These features, along with a highly extensible architecture, facilitate the visualization of various types of data such as genetic microarray data, monitoring computation grids, and other data that can be visualized as connected objects in 3D space. Future work will concentrate on further research involving interaction in virtual reality. The user will get a sense of where other users are standing, or viewing the model from. We can provide support for extra large datasets via various space partitioning techniques, such as BSP trees and octrees, and scene graph optimization techniques.

## 5  Acknowledgments

This work is supported by NSF grant ACI-0204918.